\documentclass[leqno]{article}

\usepackage{color}

\usepackage{amssymb}
\usepackage{amsfonts}
\usepackage{latexsym}
\usepackage[T1,T2A]{fontenc}
\usepackage[cp1251]{inputenc}
\usepackage[T2B]{fontenc}
\usepackage[bulgarian,russian,english]{babel}

\newtheorem {theor} {\bf Theorem}

\newtheorem{prop} {\bf Proposition}

\title { A Representation of Binary Matrices
 }
\author {Hristina Kostadinova \and  Krasimir Yordzhev}
\date {}

\begin {document}
\inputencoding{cp1251}

\maketitle

\begin{abstract}
In this article we discuss the presentation of a random binary matrix using sequence of whole nonnegative numbers. We examine some advantages and disadvantages of this presentation as an alternative of the standard presentation using
two-dimensional array. It is shown that the presentation of binary matrices using ordered n-tuples of natural numbers makes the algorithms faster and saves a lot of memory. In this work we use object-oriented programming using the syntax and the semantic of C++ programming language.
\end{abstract}

\section{Introduction.}One of the basic principles of object-oriented programming is encapsulation, which means that the client knows objects' data, properties and methods, to which events the object reacts, but it is not necessary to have the information about their realization and the algorithms used in the member functions of the corresponding class. Here we ask the following question: if there are two different classes, which describe one and the same object in mathematics or in the real world, and they have one and the same properties and methods, the question is which one of these two classes we have to choose. The answer is trivial: the class, which objects use less memory and which methods work faster, i.e. which methods use less standard computer operations.

 The present work is a continuation of the work \cite{umb2009}. Our aim is to show that the presentation of the binary matrices using ordered n-tuple of whole numbers and the bitwise operations make the realization of a better class(as it was mentioned above) compared with the standard presentation of the binary matrices using two-dimensional $n\times n$ array of whole numbers. About the definition and some examples how to use bitwise operations see \cite{davis,Kernigan,umb2009,romanov}. We recommend \cite{tan} about the object-oriented programming using C++ language in the area of the computer algebra.

We examine the set ${\cal B} =\{ 0,1\}$. ${\cal B}$ together with the operations conjunction $\&\&$, disjunction $\|$ and negation $!$ form the all known {\it boolean algebra} ${\cal B}(\&\& ,\| ,!)$, which role is very important in the computers and programming. We will use the pointed symbols for the operations in ${\cal B}(\&\& ,\| ,!)$, have in mind the semantic and the syntax of these operations in the widespread C++ programming language.

{\it Binary} matrix (or {\it boolean}, or {\it (0,1)-matrix}) is a matrix, which elements belong to the set  ${\cal B}=\{ 0,1 \}$. We denote ${\cal B}_n$ the set of all $n \times n$ square matrices.

Let $A=(a_{ij} )$ and $B=(b_{ij} )$ are matrices of ${\cal B}_n$. We consider the semantic and syntax of C++ language and the first index is 0, i.e $i,j\in \{ 0,1,2,\ldots ,n-1 \}$. We examine the following operations in ${\cal B}_n$, defined according to our aim as follows:

{\bf component conjunction}
\begin{equation}\label{1}
A \;\&\& \; B =C=(c_{ij} )
\end{equation}
where by definition for every $i,j \in \{ 0,1,2\ldots ,n-1 \} \qquad c_{ij} =a_{ij}  \;\&\& \; b_{ij}$

{\bf component disjunction}
\begin{equation}\label{2}
A \;\| \; B =C=(c_{ij} )
\end{equation}
where by definition for every $i,j \in \{ 0,1,2\ldots ,n-1 \}  \qquad c_{ij} =a_{ij}  \;\| \; b_{ij}$

{\bf component negation}
\begin{equation}\label{3}
! A =C=(c_{ij} )
\end{equation}
where by definition for every $i,j \in \{ 0,1,2\ldots ,n-1 \}  \qquad c_{ij} = ! a_{ij}$

{\bf transpose}
\begin{equation}\label{4}
t(A) =C=(c_{ij} )
\end{equation}
where by definition for every $i,j \in \{ 0,1,2\ldots ,n-1 \}  \qquad c_{ij} =a_{ji} $

{\bf logical product}
\begin{equation}\label{5}
A*B =C=(c_{ij} )
\end{equation}
where by definition for every $i,j \in \{ 0,1,2\ldots ,n-1 \} $
$$c_{ij} = \bigvee_{k=0}^{n-1} (a_{ik}  \; \&\& \; b_{kj} )=(a_{i\> 0}  \;\&\& \; b_{0\> j} )\; \|  \; (a_{i\> 1} \;\&\& \; b_{1\> j} ) \; \|  \; \cdots  \; \| \; (a_{i\> n-1}  \; \&\&  \; b_{n-1 \> j} )$$

That is the way ${\cal B}_n$ and the above-described operations $\&\&$, $\| $, $!$, $t()$ and $*$  form the algebra ${\cal B}_n (\&\& ,\| ,!,t(),*)$. Here and in the whole article the term {\it algebra} means {\it abstract algebra} considering the definition given in \cite{daintith}, and namely set equipped with various operations, assumed to satisfy some specified system of axiomatic laws. It is naturally we to put the linear order, and exactly the lexicographic order in ${\cal B}_n (\&\& ,\| ,!,t(),*)$.

We examine the following set of standard operations with integer arguments in the C++ programming language:

\begin{equation}\label{op}
 \textbf{Op}=\left\{ +,-,*,/,\% ,<<,>>,\& ,|,\wedge ,\sim , \&\& ,\| ,!, =,if  ,<,<=,>,>=,==,! \! =\right\}
\end{equation}
these operations mean addition, subtraction, multiplication, division, integer division, bitwise left shift, bitwise right shift, bitwise ''and'', bitwise ''or'', bitwise ''exclusive or'', bitwise ''negation'', conjunction, disjunction, negation, assignment, if check, and comparing. We consider that the time needed for each of these operations of the set \textbf{Op} are proportional, i.e. if  $t_1$ and $t_2$ are the times needed to execute two random operations of \textbf{Op}, then $t_1 = C t_2$, where $C$ is a const. The algorithms in this work are evaluated according to the number of the needed operations of the set \textbf{Op}.

The present work is also an appropriate example how to use bitwise operations in the object-oriented programming courses. This matter does not take enough place in the studying literature(see for example \cite{umb2009}).

\section{Two classes, which describe the algebra ${\cal B}_n (\&\& ,\| ,!,t(),*)$.}\label{sec2}

To create the first class we use the standard realization of the binary matrices: using two-dimensional $n\times n$ array of whole numbers and standard algorithms to execute the operations (\ref{1}) $\div$ (\ref{5}). Let denote this class Bn\_array.

A square binary $n\times n$ matrix, as it is described in \cite{umb2009} can be realized using ordered $n$-tuple of whole nonnegative numbers, which belong to the closed interval $[0,\; 2^n -1]$. There is one to one correspondence between the representation of the integers in decimal and in binary number system. Let denote Bn\_tuple the class which describes the algebra ${\cal B}_n (\&\& ,\| ,!,t(),*)$ using ordered $n$-tuple of whole nonnegative numbers.

These two classes have one and the same specifications (we use the terminology in \cite{azalov,todorova}), we will describe these specification using the name Bn\_X, i.e. ''X'' means ''array'', or ''tuple'' depending on the case.

Let the two classes Bn\_X have the following specification:
\begin{verbatim}
class Bn_X {
    int n;
    int *Matr;
  public:
/* constructor without parameter: */
    Bn_X();
/* constructor with parameter n pointing the row of the square matrix: */
    Bn_X (unsigned int);
/*  copy constructor: */
    Bn_X (const Bn_X &);
/* destructor: */
    ~Bn_X();
/* predefines the assignment operator: */
    Bn_X & operator = (const Bn_X &);
/* returns the size (row) of the matrix: */
    int get_n() { return n; };
/* sets value 1 to the element (i,j) of the matrix: */
    void set_1 (int,int);
/* sets value 0 to the element (i,j) of the matrix: */
    void set_0 (int,int);
/* fills row i of the matrix using integer number r
  (just for the class Bn_tuple): */
    void set_row(int,int);  // must not be included when "X" = "array"!
/* gets element (i,j) of the matrix: */
    int get_element (int,int);
/* gets the row i of the matrix
  (just for the class Bn_tuple): */
    int get_row (int);    // must not be included when "X" = "array"!
/* transposes matrix : */
    Bn_X t ();
/* predefines operators according to (1), (2), (3) и (5): */
    Bn_X operator && (Bn_X &);
    Bn_X operator || (Bn_X &);
    Bn_X operator ! ();
    Bn_X operator * (Bn_X &);
/* defines order (lexicographical) */
    int operator < (Bn_X &);
}
\end{verbatim}

When we predefine operators \&\&, || and *, if the dimensions of the two operands are not equal, then we receive the zero matrix of order the same as the first operand. But this result is not correct. We can say something more: in this case the operation is not defined, i.e. the result of the operation function is not correct and we have to be very careful in such situations. The situation is the same about the entered linear order, i.e. although the lexicographic order can be put for the words with different length, we examine by definition only the matrices of one and the same order. The result we receive when we compare the matrices with different dimensions is not correct. We give suitable massages in this situations.

Since the objects of the two classes get dynamic operation memory, using pointers and the new operator, then to work the applications, using such objects, it is necessary to predefine the operations of "the big three" \cite{azalov} - copy constructor, destructor and the assignment operator.

The constructor without parameter and the destructor are the same for the two classes Bn\_array and Bn\_tuple. Actually when we work with objects of the class Bn\_X, from a mathematical point of view it is necessary to point the dimension of the matrix and this dimension does not change. In this aspect using a constructor without parameter does not have sense and we are not interested in it. But yet we add such a constructor to make our presentation complete and to make, that "the big three" to become "the big four"\cite{azalov}.

We use the universal integer type int to save the exactness in testing the program, this type can be changed to any other whole number type.

To evaluate the effectiveness and speed of the algorithms, which use objects of the algebra ${\cal B}_n (\&\& ,\| ,!,t(),*)$ it is necessary to evaluate  the algorithms, which realize the operations $\&\&$, $\|$ $!$, $t()$, including the operation <, comparing two elements, and operation = "assignment". In that sense we describe in details just these methods, realizing the above mentioned operations. We suppose that the experienced programmer can easyly create the other methods in each of the two classes.

In the present work we predefine the operator ''<'', too. Using the same model we can predefine the other relational operators: ''<='', ''>'', ''>='', ''=='' and ''!=''. If the dimensions of the two matrices, we compare, are not equal, then the relation ''<'' is not defined and the result we get is the negative number -1.

\section{Realization of the class Bn\_array.}

When ''X''==''array'' we propose the following (\textit{standard}) realization of the examined methods in the class Bn\_array :

\begin{verbatim}
Bn_array Bn_array :: t () {
    Bn_array temp(n);
    for (int i=0; i<n; i++)
        for (int j=0; j<n; j++)
            *(temp.Matr + i*n+j) = *(Matr + j*n+i);
    return temp;
}

Bn_array Bn_array :: operator && (Bn_array &B) {
    Bn_array temp(n);
    int n2 = n*n;
    if (B.get_n() != n)
      cout<<"unallowable value of a parameter \n";
    else
        for (int p=0; p<n2; p++)
             *(temp.Matr + p) = *(this->Matr + p) && *(B.Matr + p);
    return temp;
}

Bn_array Bn_array :: operator || (Bn_array &B) {
    Bn_array temp(n);
    int n2 = n*n;
    if (B.get_n() != n)
      cout<<"unallowable value of a parameter \n";
    else
        for (int p=0; p<n2; p++)
             *(temp.Matr + p) = *(this->Matr + p) || *(B.Matr + p);
    return temp;
}

Bn_array Bn_array :: operator ! () {
    int n2 = n*n;
    for (int p=0; p<n2; p++)
        *(this->Matr + p) = *(this->Matr + p) ? 0 : 1;
    return *this;
}

Bn_array Bn_array :: operator * (Bn_array &B) {
    Bn_array temp(n);
    int c;
    if (B.get_n() != n)
      cout<<"unallowable value of a parameter \n";
      else
      for (int i=0; i<n; i++)
        for (int j=0; j<n; j++) {
          c=0;
          for (int k=0; k<n; k++) c = c || (*(this->Matr + i*n+k) && *(B.Matr +k*n+j));
          *(temp.Matr +i*n+j)=c;
        }
    return temp;
}
Bn_array& Bn_array :: operator = (const Bn_array &B) {
	int n2=n*n;
    if (B.get_n() != n)
	   cout<<"unallowable value of a parameter \n";
	else
		for (int p=0; p<n2; p++)
		   *(this->Matr + p) = *(B.Matr + p);
	return *this;
}
int Bn_array :: operator < (Bn_array &B) {
    int p = 0;
    int n2 = n*n;
    if (B.get_n() != n)
       cout<<"unallowable value of a parameter \n";
    else
         while ((*(Matr +p) == *(B.Matr +p)) && (p<n2-1) ) p++;
    if ( *(Matr +p) < *(B.Matr +p) ) return 0;
         else return 1;
}
\end{verbatim}

Analogously we can realize the remaining relational operators <=, ==, >, >=, ==, !=.

It is easy to convince that the following proposition is true:
\begin{prop} \label{p1}
For each whole positive number $n$, for the computer representation via the class Bn\_array, using the C++ programming language, of the algebra ${\cal B}_n (\&\& ,\| ,!,t(),*)$, the following propositions are true:

(i) When we use standard realization (standard predefining) of the operation functions $\&\&$, $\|$, $!$, $<$, the operation transpose and the assignment operator, each of them does $O(n^2)$ operations of the set \textbf{Op};

(ii) When we use standard realization (standard predefining) of the operation function $*$, this operation does $O(n^3)$ operations of the set \textbf{Op};

(iii) For every object of the class Bn\_array are necessary $O(n^2 )* {\rm sizeof}\> {\rm(int)}$ bytes of the operating memory;

(iv) To make initialization of the object of the class Bn\_array is done $O(n^2)$ operations of the set \textbf{Op}.

\end{prop}

We can prove the propositions (i) and (ii) as we count the number of the inner cycles and the maximal number of the iterations in each of the methods. The propositions  (iii) and (iv) are obvious.

\hfill $\Box$

\section{Representing the binary matrices using ordered $n$-tuples of whole nonnegative numbers.}

As it is shown in \cite{umb2009} there is one to one corresponding between the set of all $n\times n$  binary matrices and the set of all ordered $n$-tuples of whole numbers, which belong to the closed interval $[0,\; 2^n -1]$, based on the binary presentation of the whole numbers. This idea takes place in the realization of the class Bn\_tuple.

To create the class Bn\_tuple we propose the following methods, which realize the examined operations in the algebra ${\cal B}_n (\&\& ,\| ,!,t(),*)$. To create these methods we use bitwise operations: bitwise conjunction $\&$, bitwise disjunction $|$, bitwise exclusive ''or'' $\wedge$ and bitwise negation $\sim $, using these operations we raise the effectiveness and make the algorithms work faster.
\begin{verbatim}

Bn_tuple Bn_tuple :: t() {
	Bn_tuple temp(n);
	 int k;
	for (int i=0; i<n; i++)
		for (int j=0; j<n; j++) {
			k=get_element(i,j);
			if (k) temp.set_1(j,i);
				else temp.set_0(j,i);
		}
	return temp;
}

Bn_tuple Bn_tuple :: operator && (Bn_tuple &B) {
    Bn_tuple temp(n);
    if (B.get_n() != n)
      cout<<"unallowable value of a parameter \n";
    else
        for (int p=0; p<n; p++)
             *(temp.Matr + p) = *(this->Matr + p) & *(B.Matr + p);
    return temp;
}

Bn_tuple Bn_tuple :: operator || (Bn_tuple &B) {
    Bn_tuple temp(n);
    if (B.get_n() != n)
      cout<<"unallowable value of a parameter \n";
    else
        for (int p=0; p<n; p++)
             *(temp.Matr + p) = *(this->Matr + p) | *(B.Matr + p);
    return temp;
}

Bn_tuple Bn_tuple :: operator ! () {
	Bn_tuple temp(n);
	for (int i=0; i<n; i++) {
		 for (int j=0; j<n; j++) {
			 if ( get_element(i,j) ) temp.set_0(i,j);
				else temp.set_1(i,j);
		 }
	}
	return temp;
}

Bn_tuple Bn_tuple :: operator * (Bn_tuple &B) {
    Bn_tuple temp(n), TB(n);
    TB = t(B);
    int c, r_i, r_j;
    if (B.get_n() != n)
      cout<<"unallowable value of a parameter \n";
      else
      for (int i=0; i<n; i++)
        for (int j=0; j<n; j++) {
          r_i = this->get_row(i);
          r_j = TB.get_row(j);
          c = r_i & r_j;
          if (c==0) temp.set_0(i,j);
               else temp.set_1(i,j);
        }
    return temp;
}
Bn_tuple& Bn_tuple :: operator = (const Bn_tuple &B) {
	if (B.get_n() != n)
	   cout<<"unallowable value\n";
	else
		for (int p=0; p<n; p++)
		   *(this->Matr + p) = *(B.Matr + p);
	return *this;
}
int Bn_tuple :: operator < (Bn_tuple &B) {
	int p = 0;
	if (B.get_n() != n)
	   cout<<"unallowable value\n";
	else
		while ((get_row(p)==B.get_row(p))&&(p<n-1 )) p++;
	if (get_row(p)<B.get_row(p) ) return 1;
			  else return 0;
}
\end{verbatim}

Analogously to proposition \ref{p1} we are convinced that the following proposition is true:
\begin{prop} \label{p2}
For each whole positive number $n$, for the computer representation via the class Bn\_tuple, using the C++ programming language, of the algebra ${\cal B}_n (\&\& ,\| ,!,t(),*)$, the following propositions are true:

(i) When we use the above-described realization of the operation functions $\&\&$, $\|$, $<$ and the predefining of the assignment operator, each one of them does $O(n)$ operations of the set \textbf{Op};

(ii) The transpose and negation operation do $O(n^2)$ operations of the set \textbf{Op}.

(iii) When we use the above-described realization of the operation functions $*$, this operation does $O(n^2)$ operations of the set \textbf{Op};

(iv) For every object of the class Bn\_tuple are necessary $O(n)* {\rm sizeof}\> {\rm(int)}$ bytes of the operating memory;

(v) Initialization of the object of the class Bn\_tuple can be done using $O(n)$ operations of the set \textbf{Op}.
\end {prop}

We see that to create the class Bn\_array is easy and it is not a difficult task even for the beginning programmer, when we describe the algorithms we conform with the definitions of the corresponding operations. On the other side comparing proposition \ref{p1} with proposition \ref{p2} we are convinced that following proposition is true:

\begin{theor} \label{t1}
The algorithms using objects of the class Bn\_tuple work faster than the algorithms using objects of the class Bn\_array and they save a lot of operating memory.

\hfill $\Box$
\end{theor}

$
\begin{array}{llllll}
\mbox{Krasimir Yankov Yordzhev}   & & & & & \mbox{Hristina Aleksandrova Kostadinova}\\
\mbox{South-West University ''N. Rilsky''} & & & & & \mbox{South-West University ''N. Rilsky''}\\
\mbox{2700 Blagoevgrad, Bulgaria} & & & & & \mbox{2700 Blagoevgrad, Bulgaria}\\
\mbox{Email: yordzhev@swu.bg, iordjev@yahoo.com} & & & & & \mbox{Email: hkostadinova@gmail.com}
\end{array}
$

\end{document}